\documentclass{eas}
\usepackage{graphicx}
\begin{document}

\title{The past and future evolution of a star like Betelgeuse} 
 \runningtitle{Past and future evolution of Betelgeuse}
\author{Georges Meynet}\address{Geneva Observatory, University of Geneva, CH-1290 Versoix Switzerland}
\author{Lionel Haemmerl\'e}\sameaddress{1}
\author{Sylvia Ekstr\"om}\sameaddress{1}
\author{Cyril Georgy}\address{ENS, Lyon, France}
\author{Jos\'e Groh}\sameaddress{1}
\author{Andr\'e Maeder}\sameaddress{1}
\begin{abstract}
We discuss the physics and the evolution of a typical massive star passing through an
evolutionary stage similar to that of Betelgeuse. After a brief introduction recalling various observed
parameters of Betelgeuse, we discuss the Pre-Main-Sequence phase (PMS), the Main-Sequence (MS) phase, the
physics governing the duration of the first crossing of the HR diagram, the red supergiant stage (RSG), the post-red supergiant phases and the
final fate of solar metallicity stars with masses between 9 and 25 M$_\odot$. We examine the impact of different initial rotation and of various
prescriptions for the mass loss rates during the red supergiant phase. We show that, whatever the initial rotation rate (chosen
between 0 and 0.7$\times\upsilon_{\rm crit}$, $\upsilon_{\rm crit}$ being the surface equatorial velocity producing a centrifugal acceleration
balancing exactly the gravity) and the mass loss rates during the RSG stage (varied between a standard value and 25 times that value),
a 15 M$_\odot$ star always ends its lifetime as a RSG and explodes as a type II-P or II-L supernova.
\end{abstract}
\maketitle
\section{Introduction}
Betelgeuse (spectral type M2Iab) is one of the closest red supergiants and the subject of many studies. 
Its effective temperature is 3641$\pm$53 K  obtained by infrared interferometry (Perrin {\em et al.} \cite{2004A&A...418..675P}) and its
luminosity is around Log $L$/L$_\odot$=5.10$\pm$0.22 deduced from its bolometric luminosity and its distance which
is somewhat uncertain. The present luminosity value is deduced taking a distance of 200 pc (Harper {\em et al.} \cite{2008AJ....135.1430H},  see also the discussion in
Le Bertre {\em et al.} \cite{2012MNRAS.422.3433L}). The ratios of the abundances of nitrogen to carbon is 2.9 (in  mass fraction)
and that of nitrogen to oxygen is 0.6 according to Lambert {\em et al.} (\cite{1984ApJ...284..223L}). Typical solar values are N/C=0.3
and N/O=0.1, thus clearly Betelgeuse shows signs of the presence  of CNO processed material  on its surface (nitrogen
enhanced with respect to carbon and oxygen). The surface rotation velocity is about 5 km s$^{-1}$ (Uitenbroek {\em et al.} \cite{1998AJ....116.2501U}).
According to Le Bertre {\em et al.} (\cite{2012MNRAS.422.3433L}), Betelgeuse has been losing mass at a rate of
1.2 $\times$ 10$^{-6}$ M$_\odot$ per year for the past 80 000 years. 
The surface magnetic field is $\sim$ 1 Gauss (Auri\`ere {\em et al.} \cite{2010A&A...516L...2A}). 
In the following we shall discuss what was the past evolution of such a star, as well as what will be its future evolution.





\section{The Pre Main-Sequence phase with accretion}
\begin{figure}
\includegraphics[width=2.5in,height=3.2in]{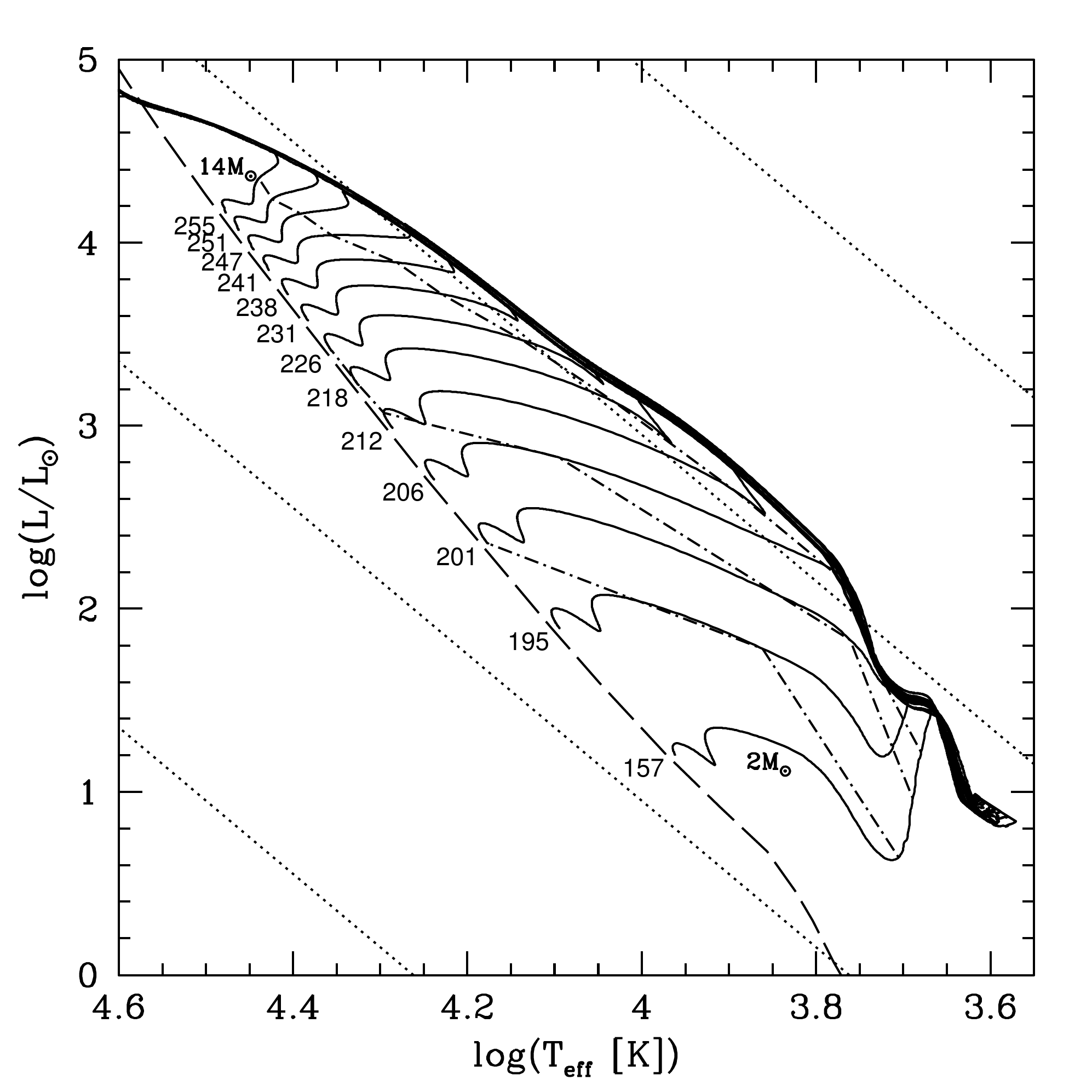}
\hfill
\includegraphics[width=2.5in,height=3.2in]{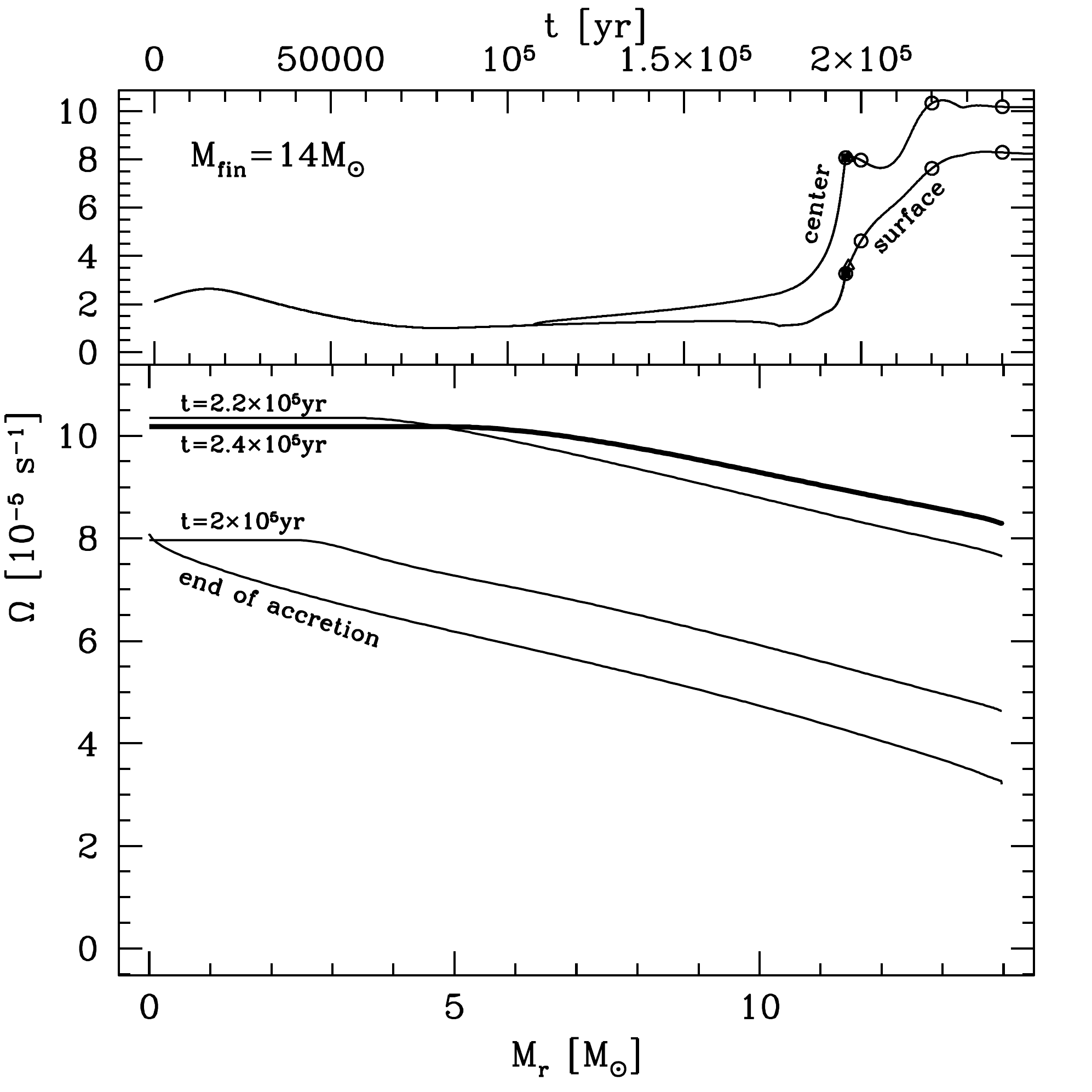}
\caption{{\it Left panel}: Grid of pre-MS models with masses at the ZAMS between 2 and 14 M$_\odot$, by steps of 1 M$_\odot$.
The dashed line is the ZAMS of Ekstr\"om {\em et al.} (\cite{2012A&A...537A.146E}),
and dotted lines are iso-radius of 0.1, 1, 10 and 100 R$_\odot$ from left to right.
The surface velocities (at the equator) on the ZAMS are indicated (in km s$^{-1}$) at the end of each track.
{\it Right panel}: Angular velocity of the model with $M_{\rm ZAMS}$ = 14 M$_\odot$.
\textit{Upper panel:} Angular velocity at the surface and in the center of the star as a function of the age.
Circles indicate the stages where the profiles of the lower panel were taken.
The filled circle indicates the end of accretion, and the triangle indicates the apparition of the convective core.
\textit{Lower panel:} Rotational profiles during the contraction phase,
at the stages indicated by a circle on the upper panel.
(The thickest line corresponds to the ZAMS.). Figures taken from Haemmerl\'e et al. (submitted to A\&A).}
\label{pms}
\end{figure}

Stars are formed from the collapse and fragmentation of large molecular clouds.
The collapse being non-homologous, hydrostatic equilibrium is first reached by a central core.
The core then grows in mass through accretion of part of the infalling material.
Today, it is believed that this scenario is occurring for low, intermediate, and also
for massive stars 
(e.g. Beech \& Mitalas \cite{1994ApJS...95..517B}, Bernasconi \& Maeder \cite{1996A&A...307..829B}, 
Yorke \& Bodenheimer \cite{2008ASPC..387..189Y}, Kuiper {\em et al.} \cite{2010ApJ...722.1556K}).

A possible pre-MS evolutionary track in the HR diagram for such a star can be seen in Fig.~\ref{pms}. The heavy line
on the right of this plot is the birthline. It corresponds to the evolutionary track of an object increasing its mass through accretion. Here the accretion law is
taken as in the work by Behrend \& Maeder (\cite{2001A&A...373..190B}). 
The computation started from a 0.7 M$_\odot$ core. Typical accretion rates are between 10$^{-5}$ and 10$^{-3}$ M$_\odot$ yr$^{-1}$. For assembling a 14 M$_\odot$ star, it takes typically  about 190 000 years. 
When, for whatever reason, accretion stops, 
the star becomes visible, hence the name of birthline given to the ``accreting'' evolutionary track. 
The  object will then evolve to reach the Zero Age Main Sequence (ZAMS). This phase lasts about 100 000 years for the 14 M$_\odot$ shown in Fig.~\ref{pms}.
We see that for such a massive star, the ``post birthline'' phase, which is also the visible phase, is quite short. 
This comes from the fact that the accretion timescale is of the same order of magnitude as
the Kelvin-Helmholtz timescale which gives the timescale for the contraction  phase. 
For those stars, both processes, accretion and slow contraction towards the ZAMS, 
take more or less the same time and since they are occurring simultaneously
when accretion stops the star is nearly already on the ZAMS. This becomes still more true for higher initial masses
and above a mass limit which depends on the accrete rate we have that the accretion timescale becomes greater than
the contraction time. In that case when contraction stops, part of the hydrogen in the core is transformed into helium (Bernasconi \& Maeder \cite{1996A&A...307..829B}). 

Haemmerl\'e {\em et al.} (submitted to A\&A) has recently revisited the scenario of formation of stars 
through accretion by studying the effects of rotation during that phase. 
In addition of accreting matter, angular momentum is accreted. 
A very simple law of angular momentum accretion has been considered: the accreted material has the same angular velocity as the surface of the accreting object,
no braking mechanism, like disk locking, is considered during the contraction phase.
On the right panel of Fig.~\ref{pms}, we can see how the internal rotation evolves during the PMS phase. When the star reaches the ZAMS
it has nearly a solid body rotation. We can compare the MS evolution obtained when the PMS phase with accretion is accounted for and when
evolution begins on the ZAMS assuming that the star rotates as  a solid body.
Very similar results are obtained provided both models on the ZAMS have the same total angular momentum.
So it means that, at least in the frame of the accretion scenarios explored by Haemmerl\'e {\em et al.}, 
the hypothesis made in many computations, {\it i.e.} to start from a solid body rotating model on the ZAMS is quite justified. 

\section{The mixing during the Main-Sequence phase}

\begin{figure}
\includegraphics[width=5in,height=2.7in]{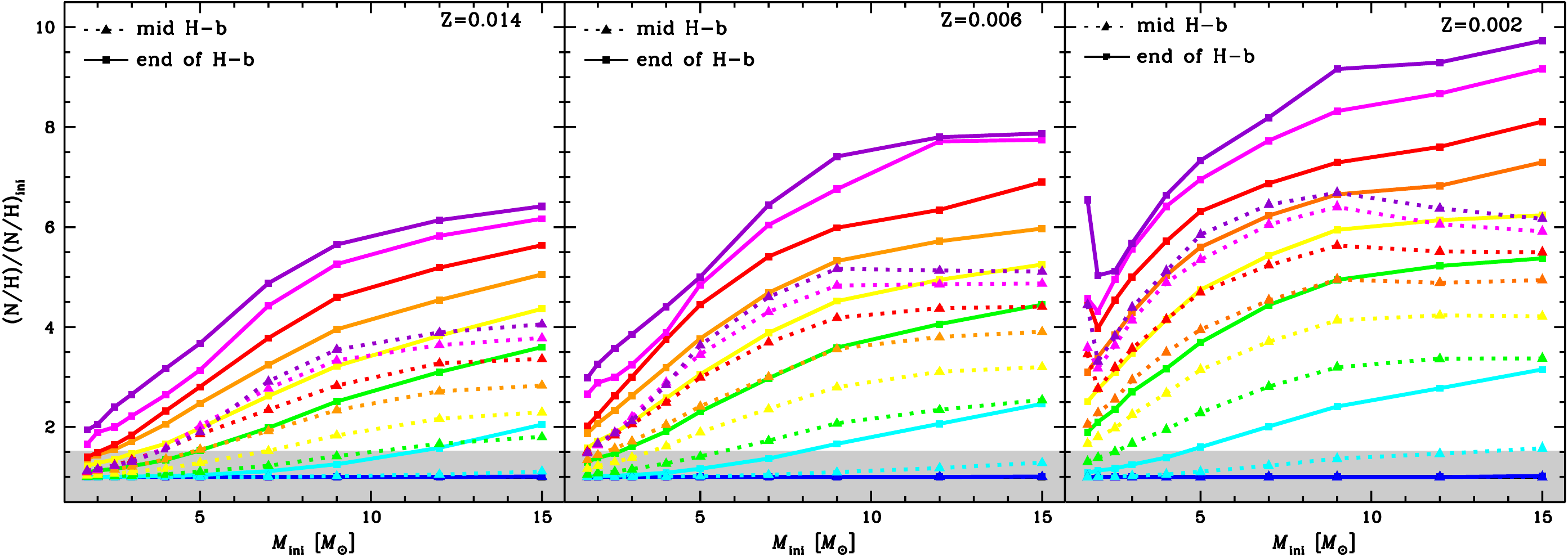}
\caption{N/H ratio at the end of the MS as a function of the initial mass for all the models. Models at $Z=0.014$ ({\it left}), $Z=0.006$ ({\it centre}), and $Z=0.002$ ({\it right}).
Figure taken from Georgy et al. (in press for A\&A). The different curves correspond to different initial velocities. From top to bottom, the angular velocity at the surface on the ZAMS corresponds to
0.95, 0.9, 0.8, 0.7, 0.6, 0.5, 0.3,  0.1 and 0 times the critical angular velocity. The curves for 0 and 0.1 are superposed and cannot be distinguished. The curves labeled {\it mid-H} correspond to the
surface abundances when the mass fraction of hydrogen at the centre is 0.35, while the curves labeled {\it end of H-b} correspond to the
surface abundances when the mass fraction of hydrogen at the centre is 0.0.}
\label{NH}
\end{figure}

In the recent years, many efforts have been made to study the impact of rotation on massive star evolution (see e.g. Langer \cite{2012ARA&A..50..107L}, Maeder \& Meynet \cite{2012RvMP...84...25M}, Chieffi \& Limongi \cite{2013ApJ...764...21C} and references therein). One of the main effects of rotation is to 
allow some elements abundant in the core to reach the surface and inversely some others, abundant in the envelope, to be injected into the convective core.
This has many consequences for the evolutionary tracks in the HR diagram, the lifetimes, the massive star populations, the nucleosynthesis, the properties
 of the final stellar collapse and of the stellar remnants.  One way of checking the physics included in these models is to observe surface
 abundances and surface velocities of B-type stars and to see whether the observations agree with the predictions of the theoretical models (see e.g. Brott {\em et al.} \cite{2011A&A...530A.116B}, Przybilla {\em et al.} \cite{2010A&A...517A..38P}). For making easier the comparison, the observed sample must be composed of 
stars as similar as possible  ({\it i.e.} same initial mass, initial composition, age, same magnetic field if any, no close companion) except for the velocity of rotation.
When such precautions are taken, general good agreement is obtained between theory and observations (Maeder {\em et al.} \cite{2009CoAst.158...72M}).
Let us note that the check of the physics of rotation is very important not only to understand single star evolution but also close binary systems with rotating components. 
  
 We show in Fig.~\ref{NH}, how the nitrogen abundance varies at the surface of different models. Without any rotation, all these models would present no enrichments,
 all the curves would be flat line in the grey zones. We note that rotational mixing increases with the initial mass, the initial rotation, with the age and when the metallicity decreases (see also e.g. Maeder \& Meynet \cite{2001A&A...373..555M}).
 A star like Betelgeuse had very likely a rotating star as a progenitor, but was it a slow or a fast rotator? We shall discuss this point further below in this paper.

\section{The first crossing of the HR diagram}

The timescale which governs the crossing of the HR diagram after the MS phase depends on when He-ignition occurs. If the ignition only occurs
when the star is in the red supergiant phase (when Log $T_{\rm eff} < 3.7$), then the timescale for the crossing
is given by the Kelvin-Helmholtz timescale and is quite short. On the contrary, when core He-ignition occurs when the star is still a blue supergiant, then 
the timescale is much longer because it is governed by the nuclear timescale. Now what are the main parameters which govern where in the HR diagram
He-ignition occurs? There is a rich literature on the subject (Renzini {\em et al.} \cite{1992ApJ...400..280R}, Stancliffe {\em et al.} \cite{2009PASA...26..203S}) and it is still a debated question. Here we discuss a few points without the ambition to be exhaustive.

In order for the envelope to expand and the star to evolve redwards in the HR diagram, some energy has to be injected in the envelope at a rate
fast enough for preventing radiation to simply evacuate it ({\it i.e.} preventing that the energy simply flows in the envelope without heating some part of it).
This ``fast'' energy injection mainly comes from the contraction of the core (part of it can come also from the H-burning shell  but the rate
of injection of energy is likely too slow and somewhat in continuity with the energy provided by the H-burning core). Among the factors that favors a weak expansion of the envelope and therefore a core He-burning phase
occurring for a significant part in the blue part of the HR diagram, one notes:
\begin{itemize}
\item a low metallicity: if we consider the extreme case of Pop III stars, most of the core He-burning phase is spent in the blue part of the HR diagram
(see e.g. Ekstr\"om {\em et al.} \cite{2008A&A...489..685E}).
The reason is the following: in absence of CNO elements, H-burning occurs at very high temperatures, so high that some 3$\alpha$ reaction takes place
too. This produces some carbon and oxygen which then trigger the activity of the CNO cycle. It means also that the central temperature
during the core H-burning phase is near the temperature for core He-burning. Thus, at the end of the MS phase, the core does not need
to contract very strongly for igniting He. Thus small amount of energy is released by core contraction and the star remains in the blue part of the HR diagram during its core He-burning phase. Only at the very end of the core He-burning phase, contraction of the core makes the star to evolve into the red supergiant stage.
Similar behavior, although less marked is expected for low (but non zero) metallicities.
\item a small helium core mass: the smaller is the helium core mass with respect to the envelope mass, the more difficult will be for the core
to release sufficient energy by contraction to uplift its envelope.
\item a weak opacity of the envelope: in case the envelope material is sufficiently transparent to radiation, then it would allow
the excess energy released by the contraction of the core to be rapidly emitted away. This would prevent the inflation of the envelope.
\item weak stellar winds: weak stellar winds allow the star to keep massive envelopes which are more difficult to inflate than
less massive ones. Note however that very strong mass losses would uncover the H-rich envelope and conversely favor a blue location in the HR diagram.
\item strong mixing: a strong mixing favors a blue location for the evolutionary tracks in the HR diagram. An illustration of this is the extreme case 
of homogeneous evolution which keeps the star in the blue part of the HR diagram (Maeder \cite{1987A&A...178..159M}). 
Fig.~\ref{mdot} shows the evolution of $T_{\rm eff}$ as a function of age during the post MS-phase for 15 M$_\odot$ solar metallicity models.
For $\upsilon_{\rm ini}/\upsilon_{\rm crit}$=0 and 0.4 , the models spend very short periods with $\log T_{\rm eff} > 4.0$, while the model
with $\upsilon_{\rm ini}/\upsilon_{\rm crit}$=0.7, in which the mixing is much stronger, spends  about one third of the post-MS phase in the blue part of the HR diagram. 
\end{itemize}
Depending on the way this first HR crossing occurs, stars enter into the RSG phase in a more or less advanced stage of their core He-burning phase.
This has consequences for the lifetime of the RSG phase.

\section{The red supergiant phase}

\begin{figure}
\includegraphics[width=2.5in,height=3.5in]{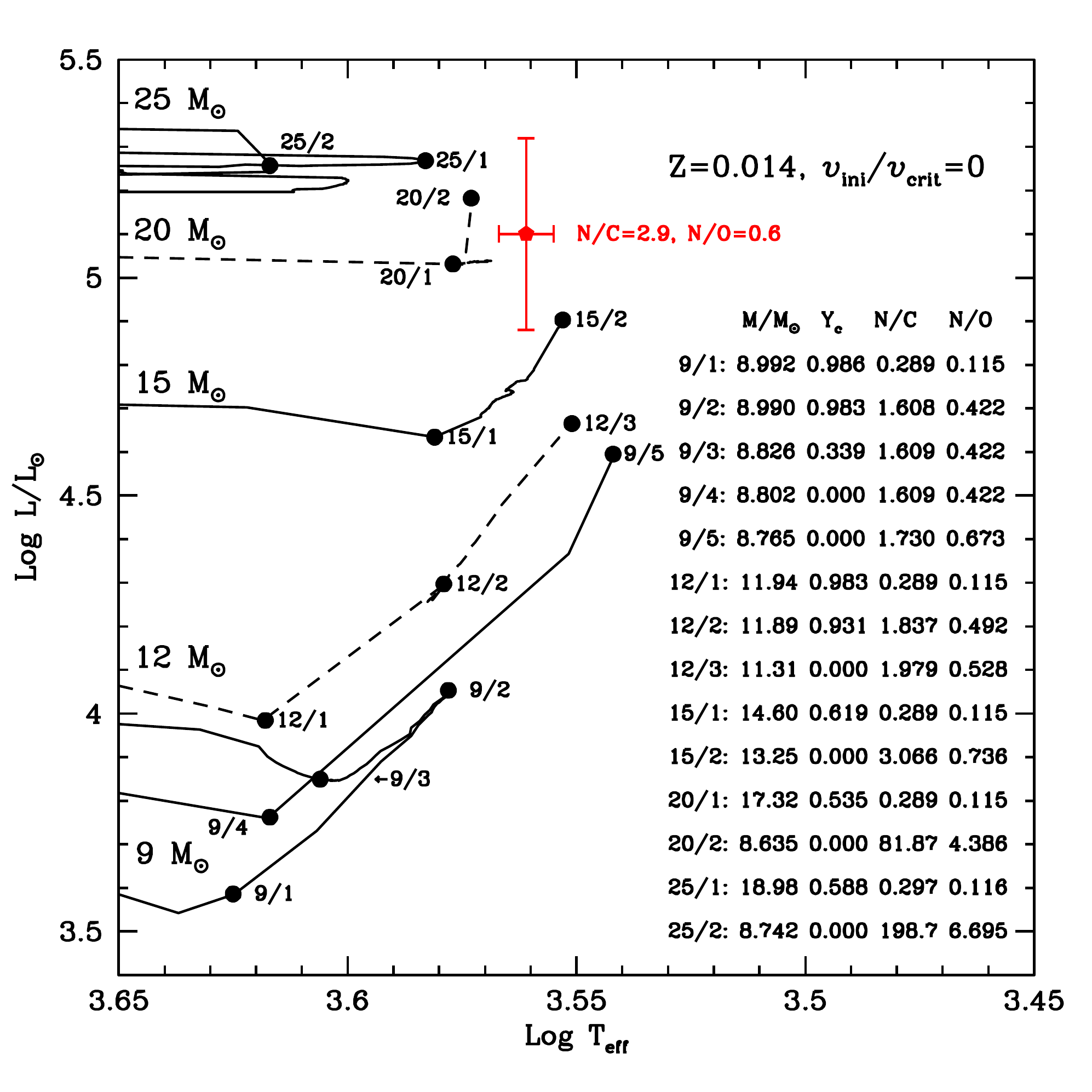}
\hfill
\includegraphics[width=2.5in,height=3.5in]{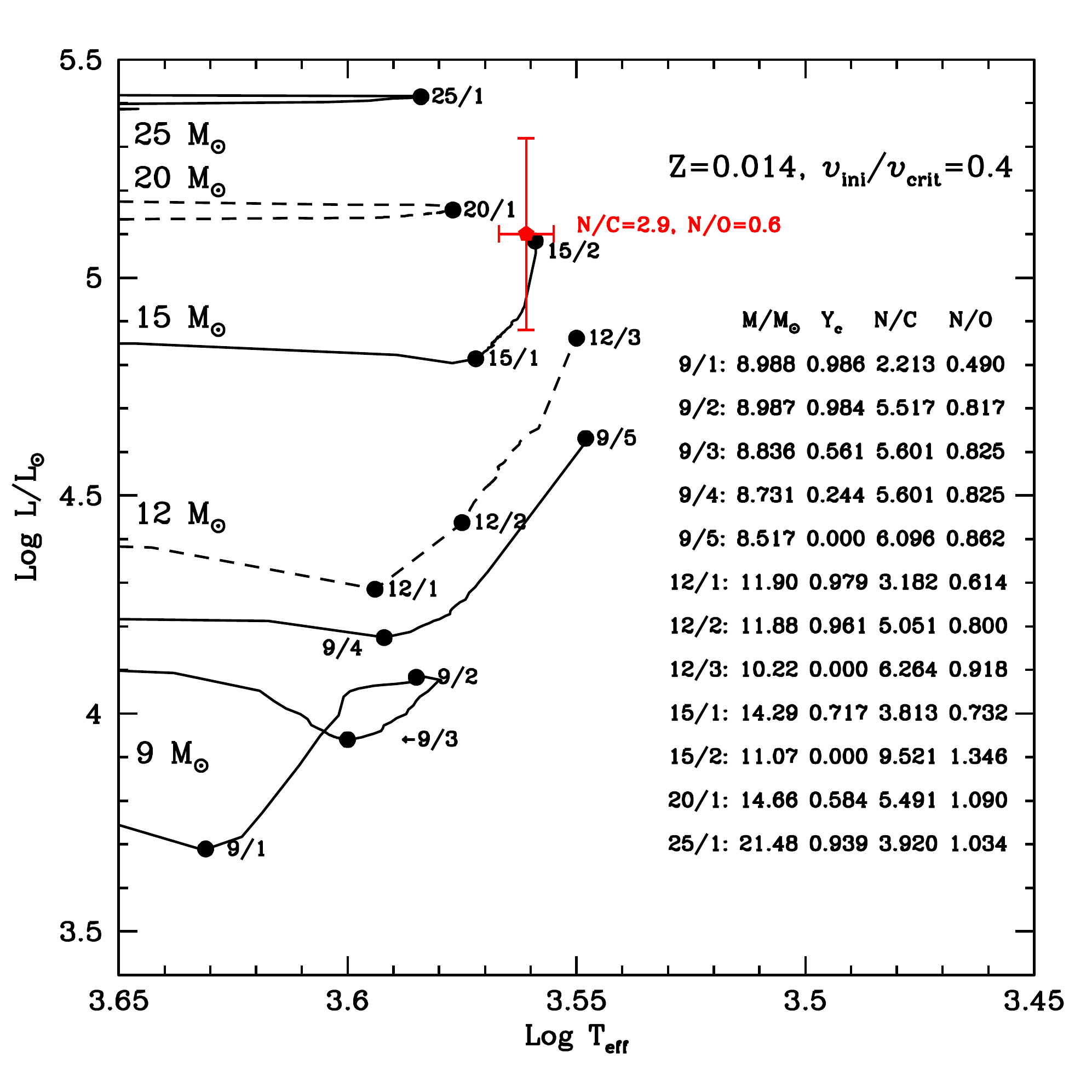}
\caption{{\it Left panel}: Evolutionary tracks for non-rotating solar-metallicity models during the red supergiant phases. For some points along the tracks, we have
indicated the actual mass of the star, the mass fraction of helium at the centre, $Y_{\rm c}$, the ratios of the abundances (in mass fraction)
at the surface of nitrogen to carbon, N/C, and of nitrogen to oxygen, N/O. The pentagon shows the position of Betelgeuse using the data reported in the introduction.
{\it Right panel}: Same as the left diagram for rotating models.}
\label{ncdhr}
\end{figure}

When the star evolves into the red supergiant phase, as a result of the deepening of the outer convective zone and of mass loss, significant changes
of the surface abundances are expected even in non-rotating models, as can be seen on the left panel of Fig.~\ref{ncdhr}.  
In this panel, one can also see that red supergiants with masses below or equal to 15 M$_\odot$ present N/C and N/O ratios which are between
initial values, 0.3 and 0.1 respectively, and maximum values equal to
$\sim$ 3.1 and 0.7 respectively. For masses equal and above 20 M$_\odot$, as a result mainly of strong mass loss
during the red supergiant phase, much higher N/C and N/O ratios can be obtained.

Looking at the right panel, one sees that when rotation is accounted for (the typical time-averaged rotation velocities of the models during the MS phase are between  190 and 220 km s$^{-1}$),
in the whole luminosity range considered, the N/C ratios (in mass fractions) are comprised between 2.2 and 9.5 and the N/O between 0.5 and 1.3.
So rotation brings three main differences: first, the minimum values of the ratios are greater than the initial ones. This is of course due to the mixing occurring during
the previous evolutionary phases. Second, the maximum values for stars with masses below 15 M$_\odot$ are slightly higher than those obtained in non-rotating models.
Third, there are no longer red supergiants in the high luminosity range presenting very high N/C and N/O ratios. This comes from the fact that the present rotating models for the 20 and 25 M$_\odot$ have very short RSG lifetimes, 
thus they do not have much time to lose important amounts of mass during that phase and thus to modify through mass loss their surface abundances during that phase.

Since observed populations are a mixture of stars with different initial velocities, one can expect some scatter of the N/C and N/O ratios observed for RSGs.
Comparing the position of Betelgeuse with the evolutionary tracks shown in Fig.~\ref{ncdhr}, we can note that rotation, through making the tracks
more luminous for a given initial mass, tends to decrease the initial mass of Betelgeuse. Non-rotating model would associate a mass around 20 M$_\odot$, while
the $\upsilon_{\rm ini}/\upsilon_{\rm crit}$=0.4 model would associate a mass around 15 M$_\odot$. Looking at Fig.~\ref{mdot}, one sees also
that the age associated to such a star would be between 11.5 and 12.5 My in case of non-rotating models, and between 15.1 and 15.7 My for
$\upsilon_{\rm ini}/\upsilon_{\rm crit}$=0.4 models. The surface N/C and N/O ratios for Betelgeuse are framed by the values of the non-rotating 20 M$_\odot$ model (see points
20/1 and 2 on the left panel of Fig.~\ref{ncdhr}) and are a little too low with respect to what is expected from rotating models (see points 15/1 and 15/2 on the right panel
of Fig.~\ref{ncdhr}), indicating that the progenitor of Betelgeuse could be a slower rotator than those shown on the right panel.

\section{What does it happen after the red supergiant phase?}
\begin{figure}
\includegraphics[width=2.7in,height=5in, angle=-90]{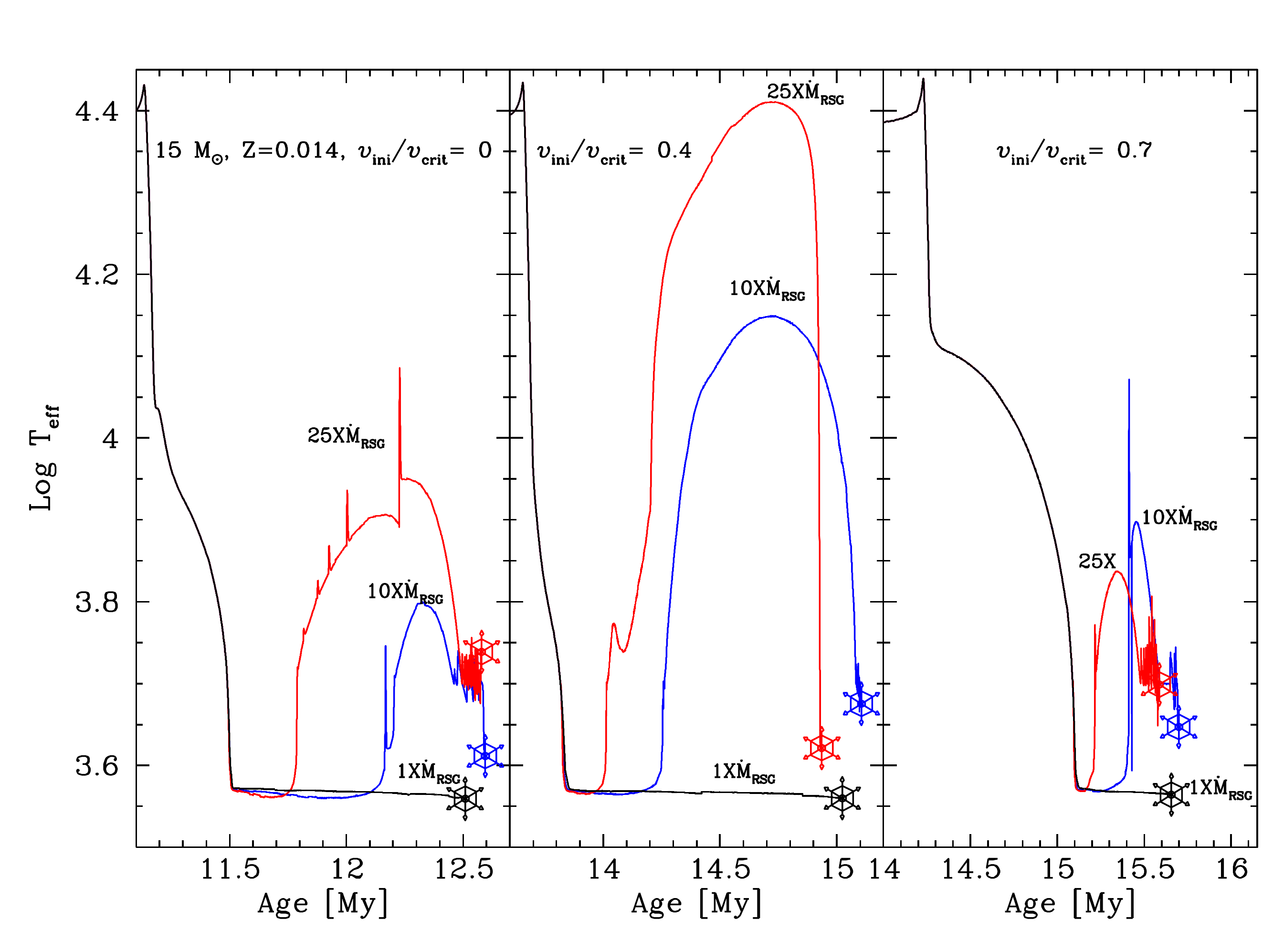}
\caption{Evolution of the effective temperature as a function of the age for 15 M$_\odot$ models at solar metallicity during the post MS phase (the small peak in Log $T_{\rm eff}$
on the left corresponds to the blue hook of the evolutionary track at the end of the MS phase) with different initial velocity and
mass loss rates during the red supergiant stage.}
\label{mdot}
\end{figure}

After the red supergiant phase, three different evolutionary scenarios may occur: 1) the star can stay in the red supergiant stage until it explodes as a type II-P supernova;
2) the star may make excursions in the blue part of the HR diagram through a blue loop evolving back to a red supergiant stage, exploding either as a type II-P or II-L supernova; 3) 
the star may end its stellar lifetime as a blue, yellow supergiant (Georgy \cite{2012A&A...538L...8G}), LBV (Groh et al. \cite{2013A&A...550L...7G}) or even a Wolf-Rayet star. Which one of these three scenarios is realized depend on mixing and mass loss. The effects of changes of mass loss rates  during the RSG phase have been studied for instance by
Vanbeveren {\em et al.} (\cite{1998NewA....3..443V}), Salasnich {\em et al.} (\cite{1999A&A...342..131S}), Yoon \& Cantiello (\cite{2010ApJ...717L..62Y}). We show in Fig.~\ref{mdot}, the tracks of 15 M$_\odot$ models at solar metallicity with different initial velocity and
mass loss rates during the red supergiant stage. The standard model (1 $\times \dot M$) has been computed with the same mass loss prescriptions as in Ekstr\"om {\em et al.} (\cite{2012A&A...537A.146E}). The two other enhanced mass loss rate models have been computed multiplying by a factor 10 and 25 the mass loss rates as long as $\log$T$_{\rm eff} < 3.7$).

Let us first comment the left panel presenting models without rotation. We see that enhanced mass loss during the RSG stage produces more extended blue loops
in effective temperatures and also increase the duration of the blue loop phase. This is expected, since mass loss enlarge the fraction of the total mass of the star
occupied by the core, and the mass fraction of the core is 
an important parameter governing the blue loop (Giannone \cite{1967ZA.....65..226G}, Lauterborn {\em et al.} \cite{1971A&A....10...97L}): the larger this mass fraction is,  the more important the loops are. The middle panel
shows the situation when some moderate rotation is accounted for. We see first that the first crossing of the HR diagram is shorter than in the non-rotating case.
This comes from the fact that for this moderate rotation rate, the cores are slightly enlarged by rotation (a factor favoring a fast redwards evolution, see Sect.~4) while the star on the whole is not too strongly mixed (a factor that would favor a blue
position in the HR diagram). We then note that the impacts of the enhanced mass losses at the red supergiant stage are
much stronger than for the non-rotating case. This comes from the fact that the cores are more massive and thus this favors at that stage a blue loop.
Finally the last panel shows the situation for quite high initial velocities. In that case, the first crossing of the HR diagram is quite long because the stars are strongly mixed.
Since the star becomes lately a red supergiant, any enhanced mass loss at that stage can only be applied during a short period. As a consequence the enhanced mass loss effects remain marginal.

These numerical simulations illustrates how mixing and mass loss can play together to enhance the blue loops or on the contrary for diminishing their importance.
In the cases studied here, all the models finish their evolution with an effective temperature below 3.7, except the non rotating 25$\times \dot M$, which ends at an 
effective temperatures slightly higher, around 3.75. 
So all these models end their life as red supergiants \footnote{Note that the models were pursued until the end of the core He-burning phase and thus there may be still some
changes in the HR position during the advanced phases. Due to the shortness of these phases, the positions indicated here should however not be too much altered
unless very strong mass losses occur.}.

\section{The final fate}
\begin{figure}
\includegraphics[width=4.5in,height=5in, angle=-90]{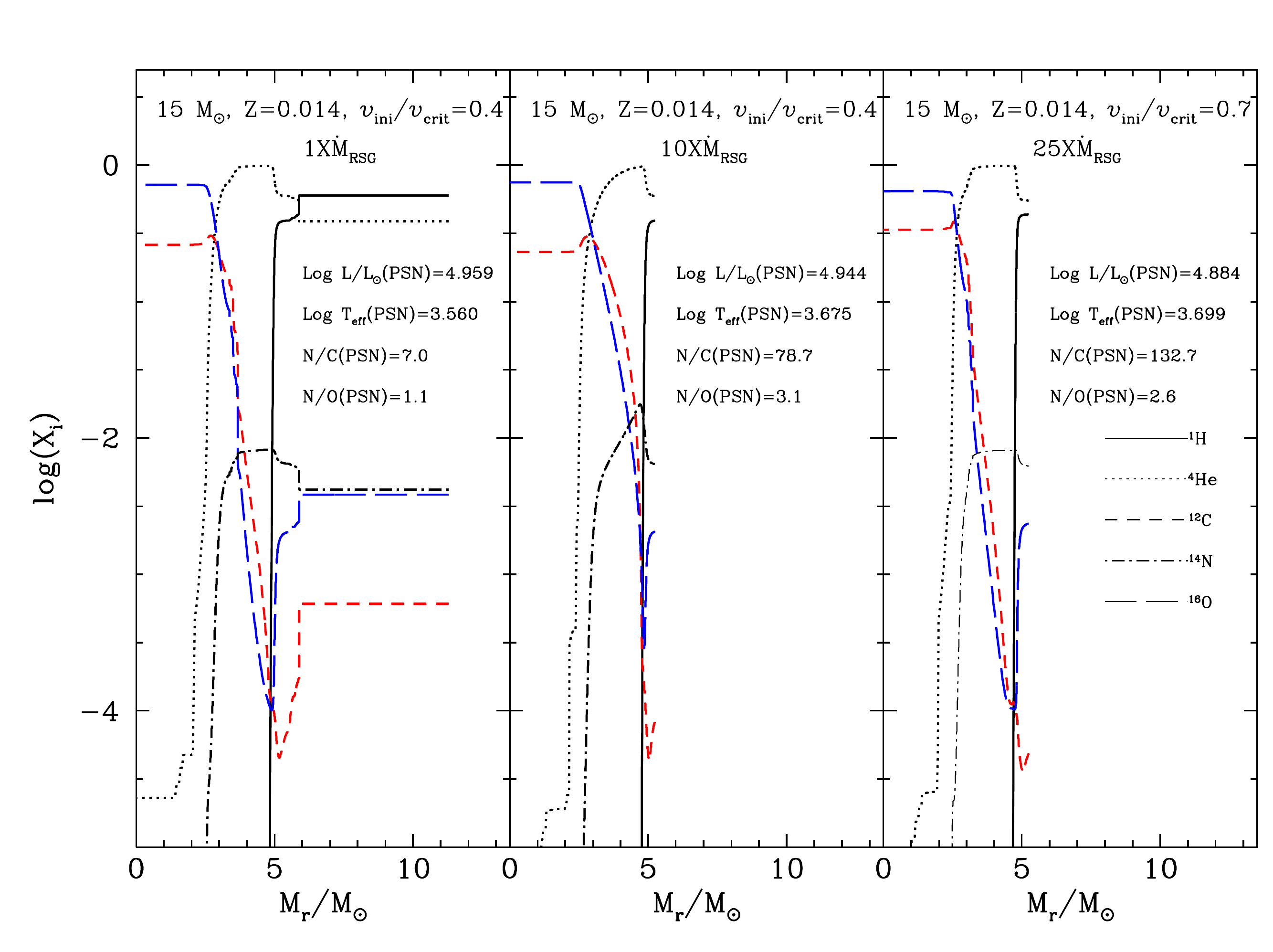}
\caption{Chemical composition of 15 M$_\odot$ solar metallicity models at the end of the core He-burning phase for models with different initial rotation
and mass loss rates during the RSG phase. The luminosity, effective temperature and the N/C, N/O ratios (in mass fraction) at the surface of the model shown are indicated. PSN is for pre-supernova}
\label{psn}
\end{figure}

The chemical structures of a few representative models for a 15 M$_\odot$ at solar metallicity are shown in Fig.~\ref{psn}. The left panel corresponds to the standard
mass loss during the RSG phase. The results for all rotational velocities are similar.
All these models end their life with a massive H-rich envelope comprising in all cases more than 5 M$_\odot$. These models will likely produce
type II-P supernovae and a neutron star. All the models with enhanced mass loss rates have structures similar to those in the middle and right panels. 
The mass has been strongly reduced, very high N/C ratio are obtained and only a very tiny H-rich layer still covers the surface. These models may explode as type II-P or II-L supernovae \footnote{Supernovae spectra
should be computed to see which type will appear. Naively from the small mass of hydrogen, one would favor a type II-L supernova.}. 

Interestingly one sees that there is no big changes between the 10 and 25$\times\dot M$ models (and this remains true for whatever the rotation is). This is because when the mass loss rates are higher, the duration of the RSG phase is shortened. Thus at the end, all the models lose more or less the same amount of mass and end their lifetimes with
quite similar structure. In Georgy {\em et al.} (\cite{2012A&A...542A..29G}), we suggested that, may be, stronger mass loss rates during the RSG phase
could produce the very low luminous WC stars observed by Sander {\em et al.} (\cite{2012A&A...540A.144S}).
The present calculations show that these WC stars cannot
be produced from 15 M$_\odot$ with enhanced RSG mass loss rates of up to 25 times the standard mass loss rates.

\section{Conclusion}
The main results of the above discussions are the following:
\begin{itemize}
\item Models of formation for a massive star like Betelgeuse through accretion indicate that
the formation time is quite short (about 300 000 years which corresponds to 1-2\% of the total lifetime).
When rotation is accounted for during the accretion process, one obtains that such a star begins its
evolution with a nearly flat rotation profile justifying the use of that approximation in most grids of stellar models
published so far.
\item The observations of the surface abundances and rotation allow to constrain the models provided
single (or non interacting) stars with similar masses, metallicities are compared at same evolutionary stages (see Maeder {\em et al.} \cite{2009CoAst.158...72M}).
\item We discussed the main parameters influencing the duration of the first crossing of the HR diagram.
\item The surface abundances given for Betelgeuse are in agreement with the predictions of models.
It is however difficult to decide whether the progenitor of Betelgeuse was slowly or moderately rotating.
\item We discussed the various evolutionary scenarios after the RSG phases and showed that
15 M$_\odot$ solar metallicity models, whatever the initial rotation between 0 and 0.7$\times \upsilon_{\rm crit}$,
are distributed between two scenarios: 1) stars remaining a RSG star until the end of their lifetime; 2) blue loops
occur but the star ends its stellar life as a red supergiant.
\item Whatever the initial rotation between 0 and 0.7$\times \upsilon_{\rm crit}$ and the mass loss rate during the RSG phase between
1 and 25$\times\dot M$, the supernova progenitor will be a RSG. There is no solution, changing the initial rotation
and/or the mass loss rate during the RSG phase producing a WR star from a single
15 M$_\odot$ star.
\end{itemize}
Beside rotation and mass loss, other parameters may influence the evolution of stars, like the initial composition, the magnetic field
and multiplicity. Let us just conclude by a question:
one can wonder whether the surface magnetic field may have an impact on the slow wind
of Betelgeuse. If we use the parameter $\eta={B^2/ 8 \pi \over \rho \upsilon^2 /2}$ giving the ratio
of the energy density in the magnetic field to that of the kinetic energy of the wind (Ud-Doula {\em et al.} \cite{2009MNRAS.392.1022U}), using the numerical values
quoted in the introduction, we obtain a value $\sim$ 1. Thus the magnetic field would just be strong enough to
have some impact on the wind. It may for instance force the matter to follow the magnetic lines and thus
extract angular momentum from the star. Such a magnetic braking may play some role in slowing down the star
not only at the surface but also in more deeper regions. Also Thirumalai \& Heyl (\cite{2012MNRAS.422.1272T}) find that the presence of a small magnetic field of about 1G
is sufficient to drive material from close to
the stellar surface up and out of its gravity well, by means of a
magneto-rotational wind.


\end{document}